\documentclass[reprint,superscriptaddress,showpacs,nofootinbib,amsmath,amssymb]{revtex4-1}
\usepackage{comment}
\usepackage{amssymb}
\usepackage{color}
\usepackage{rotating}
\usepackage{amsmath}
\usepackage{ulem}
\usepackage{overpic}
\usepackage{graphicx}
\usepackage{dcolumn}
\usepackage{bm}
\usepackage{chngpage}
\usepackage{multirow}
\usepackage{slashed}
\usepackage{indentfirst}
\usepackage{booktabs}
\usepackage{amssymb,amsfonts}
\usepackage{amsmath}
\usepackage{color}
\usepackage[colorlinks,citecolor=blue,linkcolor=blue,hypertex,breaklinks=true]{hyperref}

\begin{document}
\title{Implication of shell quenching in the scandium isotopes around $N=20$}
\author{Rong An}
\affiliation{School of Physics, Ningxia University, Yinchuan 750021, China}
\affiliation{Guangxi Key Laboratory of Nuclear Physics and Technology, Guangxi Normal University, Guilin, 541004, China}
\affiliation{Key Laboratory of Beam Technology of Ministry of Education, School of Physics and Astronomy, Beijing Normal University, Beijing 100875, China}

\author{Xiang Jiang}
\affiliation{College of Physics and Optoelectronic Engineering, Shenzhen University, Shenzhen 518060, China}

\author{Na Tang}
\affiliation{School of Physics, Ningxia University, Yinchuan 750021, China}

\author{Li-Gang Cao}
\email[Corresponding author:]{caolg@bnu.edu.cn}
\affiliation{Key Laboratory of Beam Technology of Ministry of Education, School of Physics and Astronomy, Beijing Normal University, Beijing 100875, China}
\affiliation{Key Laboratory of Beam Technology of Ministry of Education, Institute of Radiation Technology, Beijing Academy of Science and Technology, Beijing 100875, China}

\author{Feng-Shou Zhang}
\email[Corresponding author:]{fszhang@bnu.edu.cn}
\affiliation{Key Laboratory of Beam Technology of Ministry of Education, School of Physics and Astronomy, Beijing Normal University, Beijing 100875, China}
\affiliation{Key Laboratory of Beam Technology of Ministry of Education, Institute of Radiation Technology, Beijing Academy of Science and Technology, Beijing 100875, China}
\affiliation{Center of Theoretical Nuclear Physics, National Laboratory of Heavy Ion Accelerator of Lanzhou, Lanzhou 730000, China}

\date{\today}

\begin{abstract}
  Shell closure structures are commonly observed phenomena associated with nuclear charge radii throughout the nuclide chart.
  Inspired by recent studies demonstrating that the abrupt change can be clearly observed in the charge radii of the scandium isotopic chain across the neutron number $N=20$, we further review the underlying mechanism of the enlarged charge radii for $^{42}$Sc based on the covariant density functional theory. The pairing correlations are tackled by solving the state-dependent Bardeen-Cooper-Schrieffer equations. Meanwhile, the neutron-proton correlation around the Fermi surface derived from the simultaneously unpaired proton and neutron is appropriately considered in describing the systematic evolution of nuclear charge radii. The calculated results suggest that the abrupt increase in charge radii across the $N=20$ shell closure seems to be improved along the scandium isotopic chain if the strong neutron-proton correlation is properly included.
\end{abstract}


\maketitle

\section{Introduction}\label{intro}
For self-bound complex nuclei systems, quantum shell closure effects are generally observed in various physical quantities, such as $\alpha$-decay properties~\cite{PhysRevC.98.034312,PhysRevLett.122.192503,PhysRevC.108.064303,PhysRevC.109.054310,PhysRevC.109.L011301}, nuclear mass evolution~\cite{XIA20181,mass2020,ZHANG2022101488,ADNDT158_101661}, and the systematic trend of changes of nuclear charge radii~\cite{PhysRevC.88.011301,PhysRevC.100.044310,PhysRevC.104.064313,PhysRevC.105.014325}.
The fully filled shell results in the relatively stable properties of a nucleus with respect to the adjacent counterparts.
Particularly, along a long isotopic chain, the shrunken trend in nuclear charge radii (being the kink phenomenon) can be evidently observed at the fully filled neutron numbers $N=28$, $50$, $82$, and $126$~\cite{ANGELI201369,LI2021101440}.
This leads to the local minima of proton radii along a long isotopic chain. Meanwhile, imprints of the shell closures at the neutron numbers $N=14$ and $N=16$ can also be recognized from the proton density distributions~\cite{BAGCHI2019251,PhysRevC.102.051303,PhysRevLett.129.142502}.
This means that the signature of neutron magicity could be identified from the local variations of nuclear charge radii.

Charge radii of neutron-rich isotopes beyond the neutron-closed shell $N=28$ appear to increase with astonishingly similar slopes~\cite{GarciaRuiz:2019cog,PhysRevC.105.L021303}.
Across the traditional neutron magic numbers $N=20$, the abrupt change of nuclear charge radii is unexpectedly disappeared in the calcium isotopes. The same scenarios can also be encountered in the neighboring Ar~\cite{KLEIN19961} and K~\cite{TOUCHARD1982169,PhysRevLett.79.375,PhysRevC.92.014305} isotopic chains, namely only smooth variations have been almost presented in the charge radii.
The apparent disappearance of the rapid increase in nuclear charge radii poses a long-standing challenge to nuclear many-body theory across the traditional neutron number $N=20$.
In recent study, a potentially reduced charge radius of $^{32}$Al seems to support an effect of the $N=20$ shell closure from the aspect of nuclei size~\cite{PhysRevC.103.014318}.
The latest study shows that charge radii of neutron deficient $^{40}$Sc (3.514$\pm$0.025 fm) and $^{41}$Sc (3.503$\pm$0.020 fm) isotopes have been detected using collinear laser spectroscopy technique~\cite{PhysRevLett.131.102501}.
Combining the existing charge radius of $^{42}$Sc (3.557$\pm$0.014 fm) in the literature~\cite{Avgoulea_2011}, the abrupt increase of charge radii can be surprisingly observed from $^{41}$Sc to $^{42}$Sc.
This suggests that a pronounced kink phenomenon at the neutron number $N=20$ has been verified from the aspect of nuclear charge radii in the scandium isotopes.

The undertaken efforts have been devoted to describing the trend of changes of nuclear charge radii, such as the generally applied empirical formulas ruled by the $A^{1/3}$ or $Z^{1/3}$ law~\cite{Bohr1969,Zhang:2001nt}, relativistic~\cite{PhysRevC.82.054319,geng2003} and non-relativistic~\cite{PhysRevC.82.035804,PhysRevLett.102.242501,PhysRevC.95.064328} energy density functionals, and the machine learning approaches~\cite{Utama_2016,PhysRevC.105.014308,zhengzhenhua2022,PhysRevC.108.034315,PhysRevC.110.014316,PhysRevC.110.014308,new2024}.
Here, it is obviously mentioned that the microscopic mechanisms cannot be properly captured by the empirical formulas.
As suggested in Ref.~\cite{PhysRevLett.131.102501}, the charge radius of $^{41}$Sc is significantly below the value of $^{42}$Sc,
and this results in the pronounced kink structure at the neutron number $N=20$ along the scandium isotopic chain.
However, both of the $ab$ $initio$ and density functional theory (DFT) models cannot reproduce the intriguingly increasing pattern of charge radii for neutron-deficient scandium isotopes.
Nevertheless, as demonstrated in Refs.~\cite{PhysRevC.102.024307,Inakura:2024mri}, the local variations of nuclear charge radii cannot be reproduced well at the mean-field level.

The discontinuous variations in nuclear charge radii, such as the shell quenching phenomena, can be described well by incorporating the Casten factor, in which the neutron-proton correlations are derived from the valence neutrons and protons~\cite{PhysRevLett.58.658,Angeli_1991,Sheng2015,Xian2024}. The similar approach which considers the neutron-proton correlations around Fermi surface in the root-mean-square (rms) charge radii formula has been proposed to describe the fine structure of nuclear size through the covariant density function theory (CDFT) model.
This modified method can reproduce the anomalous behaviors of charge radii along an long isotopic family, but the odd-even staggering (OES) effect is profoundly overestimated along odd-$Z$ isotopic chains~\cite{An_202201,An_202202}.
To melt this tension, the neutron-proton correlation derived from the simultaneously unpaired neutron and proton around Fermi surface has been incorporated into the rms charge radii formula~\cite{PhysRevC.109.064302}.
Although the correlation between the simultaneously unpaired neutron and proton has been considered, the calculated charge radius of $^{42}$Sc is obviously underestimated.
This seems to suggest that the neutron-proton correlation cannot be captured adequately for the unpaired nucleons yet.
Therefore, the underlying mechanism should be clarified for our deeper understanding.

For the isotope $^{42}$Sc, the lastly unpaired proton and neutron occupy the 1f$_{7/2}$ shell with respect to the relatively stable core $^{40}$Ca.
The blocking approximation is just employed to tackle the unpaired nucleons~\cite{PhysRevC.109.064302}. For these nucleons with time-reversal symmetry breaking, the unpaired single particle occupations should be tackled appropriately.
In order to further review the underlying mechanism for the enlarged charge radii from $^{41}$Sc to $^{42}$Sc, the CDFT model is still employed in this work.
The pairing correlations are properly tackled by solving the state-dependent Bardeen-Cooper-Schrieffer equations~\cite{geng2003}.
The quadrupole deformation and the neutron-proton pairs correlation around Fermi surface are also incorporated into our discussion.
The influence of different occupations of the proton and neutron single particle levels on the charge radius of $^{42}$Sc is also inspected for the simultaneously unpaired neutron and proton in this literature.

The structure of the paper is the following. In Section~\ref{Sec3},
the theoretical framework is briefly presented. In Section~\ref{Sec4},
the numerical results and discussion are provided. Finally, a
summary is given in Section~\ref{Sec5}.
\section{Theoretical framework}\label{Sec3}
The covariant density functional theory (CDFT) model has made greatly successful in describing various physical phenomena~\cite{PhysRevLett.77.3963,VRETENAR2005101,Cao:2003yn,PhysRevC.67.034312,PhysRevC.68.034323,PhysRevC.69.054303,MENG2006470,PhysRevC.82.011301,PhysRevC.92.024324,SUN2018530,An_2020,SUN2020122011,Wang_2022,RONG2023137896,PhysRevC.108.054314,PhysRevC.108.L041301,PhysRevC.109.024321}.
For nonlinear self-consistent Lagrangian density, the effective interactions among nucleons are provided by the exchange of $\sigma$, $\omega$ and $\rho$ mesons. The electromagnetic interaction is offered by photons naturally. The effective Lagrangian density is recalled as follows~\cite{RING199777}:
\begin{eqnarray}
\mathcal{L}&=&\bar{\psi}[i\gamma^\mu\partial_\mu-M-g_\sigma\sigma
-\gamma^\mu(g_\omega\omega_\mu+g_\rho\vec
{\tau}\cdotp\vec{\rho}_{\mu}+eA_\mu)]\psi\nonumber\\
&+&\frac{1}{2}\partial^\mu\sigma\partial_\mu\sigma-\frac{1}{2}m_\sigma^2\sigma^2
-\frac{1}{3}g_{2}\sigma^{3}-\frac{1}{4}g_{3}\sigma^{4}\nonumber\\
&-&\frac{1}{4}\Omega^{\mu\nu}\Omega_{\mu\nu}+\frac{1}{2}m_{\omega}^2\omega_\mu\omega^\mu
+\frac{1}{4}c_{3}(\omega^{\mu}\omega_{\mu})^{2}-\frac{1}{4}\vec{R}_{\mu\nu}\cdotp\vec{R}^{\mu\nu}\nonumber\\
&+&\frac{1}{2}m_\rho^2\vec{\rho}^\mu\cdotp\vec{\rho}_\mu
+\frac{1}{4}d_{3}(\vec{\rho}^{\mu}\vec{\rho}_{\mu})^{2}-\frac{1}{4}F^{\mu\nu}F_{\mu\nu},
\end{eqnarray}
where $M$ is the mass of nucleon and $m_{\sigma}$, $m_{\omega}$, and $m_{\rho}$, are the masses of the $\sigma$, $\omega$ and $\rho$ mesons, respectively. Here $g_{\sigma}$, $g_{\omega}$, $g_{\rho}$, $g_{2}$, $g_{3}$, $c_{3}$, $d_{3}$ and $e^{2}/4\pi$ correspond to the coupling constants for $\sigma$, $\omega$, $\rho$ mesons and photon, respectively.

The Dirac equation with effective fields $S(\mathbf{r})$ and $V(\mathbf{r})$ and Klein-Gordon equations with various mesons sources are derived through variational principle~\cite{RING199777}.
In order to capture the ground-state properties of finite nuclei, the quadrupole deformation parameter $\beta_{20}$ becomes constrained in the self-consistently iterative process.
Therefore, the Hamiltonian formalism can be rewritten as follows:
\begin{eqnarray}
\mathcal{H}=\bf{\alpha}\cdot\mathbf{p}+V(\mathbf{r})+\beta{[M+S(\mathbf{r})]}-\lambda\mathbf{Q},
\end{eqnarray}
where $\lambda$ is the spring constant, and $\mathbf{Q}$ is the intrinsic quadrupole moment operator.
Those values of the corresponding quadrupole deformation parameter $\beta_{20}$ are changed from $-0.25$ to $0.25$ with the interval range of $0.01$.

To account for the implications of the observed odd-even oscillations of nuclear charge radii along odd-$Z$ isotopic chains, the further modified mean-square charge radii formula has been proposed as follows (in units of fm$^2$)~\cite{PhysRevC.109.064302},
\begin{eqnarray}\label{cp2}
R_{\mathrm{ch}}^{2}=\langle{r_{\mathrm{p}}^{2}}\rangle+0.7056~\mathrm{fm^{2}}+\frac{a_{0}}{\sqrt{A}}\Delta{\mathcal{D}}~\mathrm{fm^{2}}+\frac{\delta}{\sqrt{A}}~\mathrm{fm^{2}}.
\end{eqnarray}
The first term represents the charge distribution of point-like protons and the second term is due to the finite size of protons~\cite{Gambhir:1989mp}.
Here, the quantity of the proton radius takes the values about $0.84$ fm~\cite{RevModPhys.93.025010,PhysRevLett.128.052002}.
As shown in Eq.~(\ref{cp2}), the quantity of $|\Delta\mathcal{D}|$ is associated with the difference of the Cooper pairs condensation between the neutrons and protons~\cite{PhysRevC.76.011302}.
The effective force in the covariant density functional theory is chosen to be the parameter set NL3~\cite{PhysRevC.55.540}.
The last term in this expression just represents the neutron-proton correlation deriving from the simultaneously unpaired neutron and proton.
This means the quantity of $\delta$ equals to zero for even-even, even-odd, and odd-even nuclei.
The values of $a_{0}=0.561$ and $\delta=0.355$ are calibrated by fitting the odd-even oscillation and the inverted parabolic-like shape of charge radii along potassium and calcium isotopes under effective force NL3~\cite{PhysRevC.109.064302}.
The pairing strength is determined through the empirical odd-even mass staggering~\cite{Bender:2000xk}.
As shown in Ref.~\cite{PhysRevC.109.064302}, the pairing strength is chosen to $V_{0}=350$ MeV fm$^{3}$ for effective force NL3 set.

\section{Results and discussion}\label{Sec4}
\begin{figure*}
\resizebox{1.10\textwidth}{!}{%
  \includegraphics{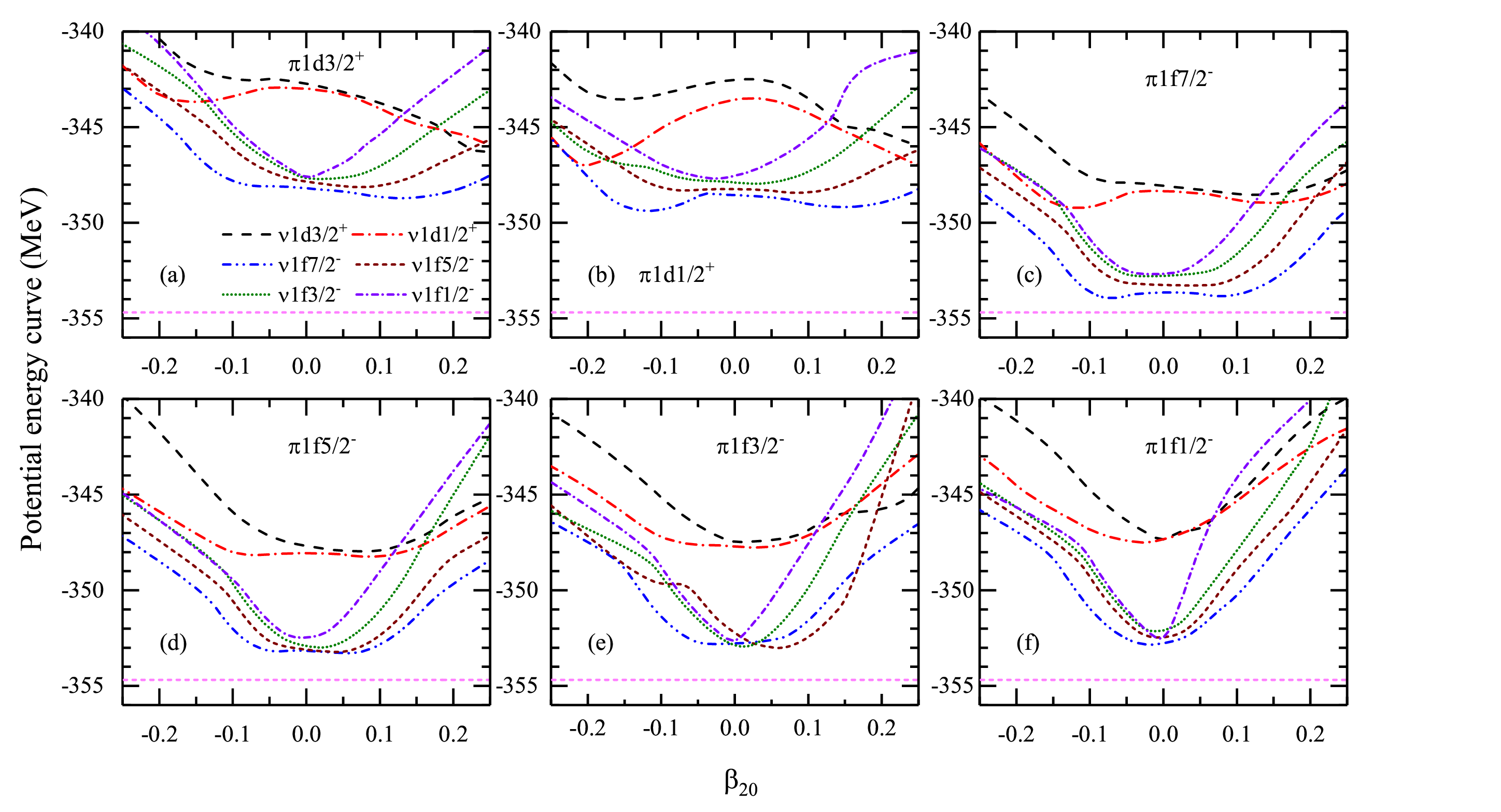}
}
\caption{(Color online) Potential energy curve (PEC) of $^{42}$Sc as functions of the quadrupole deformation $\beta_{20}$ vs different configurations mixing of single particle orbits occupied by the last unpaired nucleons. $\pi$ and $\nu$ denote the last unpaired proton and neutron, respectively, where the components of the angular momentum of the 1d$_{3/2}$ ($j_{z}=3/2^{+}$, $j_{z}=1/2^{+}$) and 1f$_{7/2}$ ($j_{z}=7/2^{-}$, $j_{z}=5/2^{-}$, $j_{z}=3/2^{-}$, $j_{z}=1/2^{-}$) subshells projected on the z-axis are used to denote the occupied states. Experimental datum is taken from Ref.~\cite{mass2020}, which is represented by the horizontal short line.}
\label{fig1}       
\end{figure*}
The signatures of neutron magic numbers can be reflected through the binding energies, nuclear charge radii, excitation energies,and transition probabilities~\cite{GarciaRuiz:2019cog,SORLIN2008602,Angeli_2015}.
As strongly correlated many-body system, the rapid raise of nuclear charge radii along a long isotopic chain results from different mechanism, such as
the shape-phase transition~\cite{RevModPhys.83.1467,PhysRevLett.117.172502,An2023035301}, and the shell closure effect~\cite{PhysRevC.88.011301,PhysRevC.100.044310,PhysRevC.104.064313,PhysRevC.105.014325}.
The isotopes with neutron magic numbers persist relatively stable binding energy, and the proton density distributions exhibit rather smaller variations in comparison with the neighboring counterparts.
In this work, the calculated result suggests that the quadrupole deformation $\beta_{20}$ for $^{42}$Sc is about $-0.07$, namely the almost spherical shape.
This is in accord with the value shown in Ref.~\cite{MOLLER20161}.
For $^{41}$Sc, the spherical shape is also presented~\cite{PhysRevC.109.064302}.
This means that the influence coming from the quadrupole deformation is not adequate yet in reproducing the abruptly increasing trend of charge radii from $^{41}$Sc to $^{42}$Sc isotopes.
The rapid increase of charge radius can be significantly observed across the $^{41}$Sc isotope~\cite{PhysRevLett.131.102501}. However, this profound kink structure at the $N=20$ fully filled-shell is absent in the neighboring Ca, K, and Ar isotopic chains.

\begin{figure*}[!htb]
\resizebox{1.10\textwidth}{!}{%
  \includegraphics{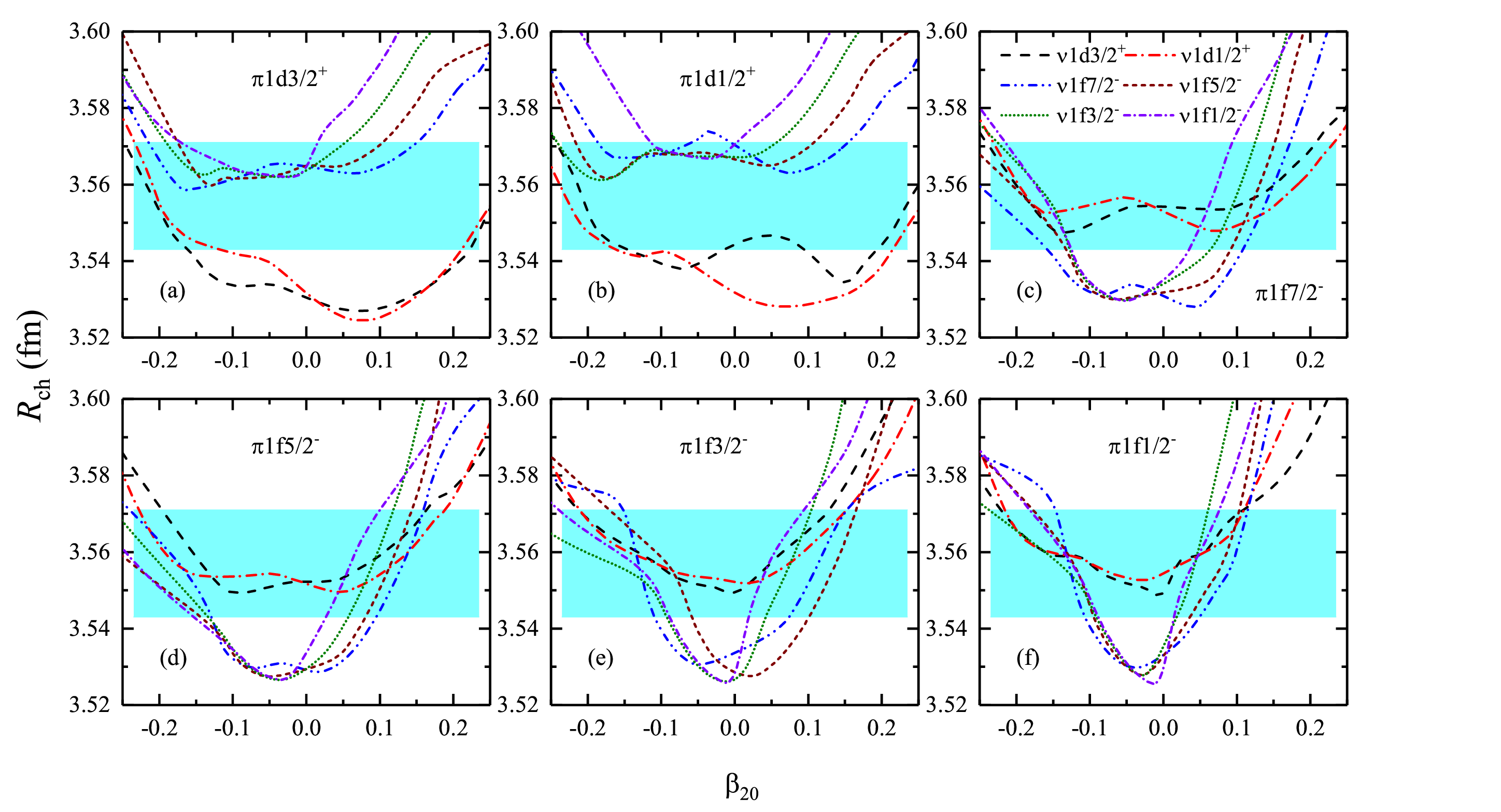}
}
\caption{(Color online) Same as Fig.~\ref{fig1} but for the root-mean-square (rms) charge radii. Experimental data are taken from Refs.~\cite{PhysRevLett.131.102501,Avgoulea_2011}, where the light blue region represents the systematic error band. }
\label{fig2}
\end{figure*}

Along Sc isotopic family, the time-reverse symmetry is broken by the last unpaired proton.
In general, it is mentioned that the blocking approximation is employed in tackling the occupation of the last unpaired proton and neutron.
The occupation numbers of unpaired neutrons and protons are determined by adjusting them to minimize the nuclear binding energy.
Therefore, the blocking treatment of the last unpaired neutron and proton is further inspected in the following discussion.
Meanwhile, it is mentioned that the quadrupole deformation has an influence on the determination of nuclear charge radii~\cite{PhysRevC.108.024310}. Therefore, the potential energy curves (PEC) of $^{42}$Sc would be depicted for various blocking states.
Shown in Fig.~\ref{fig1}, the PEC of $^{42}$Sc as a function of the quadrupole deformation $\beta_{20}$ is plotted. For the convenience of discussion, the projection of the angular momentum on the z-axis in the 1d$_{3/2}$ ($j_{z}=3/2^{+}$, $j_{z}=1/2^{+}$) and 1f$_{7/2}$ ($j_{z}=7/2^{-}$, $j_{z}=5/2^{-}$, $j_{z}=3/2^{-}$, $j_{z}=1/2^{-}$) shells are employed to declare the occupied orbits for the last unpaired proton ($\pi$) and neutron ($\nu$), in which the superscripts $+$ and $-$ represent the parity.
The last unpaired neutron ($\nu$) locates sequentially at the single particle levels from the 1d$_{3/2}$ to 1f$_{7/2}$ subshells.
As shown in Figs.~\ref{fig1}(a) and (b), the 1d$_{3/2}$ level is occupied in order by the last unpaired proton.
One can find that the local minimum in PEC of $^{42}$Sc diverges heavily from the experimental datum.
While for $\nu$1f$7/2^{-}$ ($j_{z}=7/2^{-}$), the calculated result is relatively close to the experimental datum, but about $7.0$ MeV deviation can be obtained.

Shown in Fig.~\ref{fig1}(c), the last unpaired proton occupies the $j_{z}=7/2^{-}$ level.
The largest binding energy of $^{42}$Sc can be obtained if the last unpaired neutron occupies the $j_{z}=7/2^{-}$ level.
Here one can mention that the quadrupole deformation locates at the value of $\beta_{20}=-0.07$, namely the almost spherical shape.
As shown in Figs.~\ref{fig1}(d), (e), and (f), the levels $\pi$1f$5/2^{-}$, $\pi$1f$3/2^{-}$, and $\pi$1f$1/2^{-}$ are sequentially occupied by the last unpaired proton.
It is found that the local minimum energies are almost similar or equivalent for these cases.
This is due to the fact that quadrupole deformation values are extremely close to each other for the corresponding configuration mixing of the last unpaired single particle levels.
The 1f$7/2^{-}$ level occupied by the simultaneously unpaired neutron and proton gives raise to the relatively stable ground-state property of $^{42}$Sc.
Although the configuration combinations through various levels $\pi$1f$5/2^{-}$, $\pi$1f$3/2^{-}$, and $\pi$1f$1/2^{-}$ give the similar results, the slight deviations can also be distinguished in comparison with the case that the $\pi$1f$7/2^{-}$ level occupied by the simultaneously unpaired nucleons.

To facilitate the quantitative comparison, the root-mean-square (rms) charge radii of $^{42}$Sc are also plotted in Fig.~\ref{fig2} as a function of quadrupole deformation $\beta_{20}$ under various configuration mixing occupations of the simultaneously unpaired neutron and proton.
The calculated results can cover the uncertainty range of charge radius of $^{42}$Sc when the last unpaired proton and neutron occupy the 1d$_{3/2}$ and 1f$_{7/2}$ subshells, respectively.
However, for these cases, the ground-state energies are deviated heavily from the experiment, namely the loosely bound states are expected.
The same scenarios can also be encountered in the configuration where the unpaired proton occupies the 1f$_{7/2}$ subshell and the 1d$_{3/2}$ subshell is occupied by the unpaired neutron.
For the configurations mixing with various levels in the 1f$_{7/2}$ subshell, each of these occupied combinations gives the lower value than the experimental one.
This means that the experimental value cannot be reproduced well through various configurations mixing of the last unpaired neutron and proton.

The signature of local variations can be manifested in nuclear charge radii.
In order to further inspect these discontinuous variations of charge radii along a specific isotopic family, the three-point formula has been recalled as follows~\cite{PhysRevC.95.064328,PhysRevC.102.024307},
\begin{small}
\begin{eqnarray}\label{oef1}
\Delta_{r}(N,Z)=\frac{1}{2}[2R(N,Z)-R(N-1,Z)-R(N+1,Z)],
\end{eqnarray}
\end{small}
where $R(N,Z)$ is rms charge radius for a nucleus with neutron number $N$ and proton number $Z$.
As discussed above, the charge radius of $^{42}$Sc cannot be reproduced well by blocking various configurations mixing of the simultaneously unpaired neutron and proton.
Meanwhile, combining various occupations of the unpaired single particle levels in our model, the quadrupole deformation parameters $\beta_{20}$ are almost spherical. Hence the abrupt raise of charge radii due to the shape-phase transition is also excluded in our discussion.

Considering the neutron-proton correlation around Fermi surface on the simultaneously unpaired neutron and proton, the odd-even staggering (OES) behavior of charge radii can be characterized well along odd-$Z$ isotopic chains.
In particular, the OES of charge radii along potassium and copper isotopes can be reproduced well.
However, in the scandium isotopic family, the abrupt increase of charge radii from $^{41}$Sc to $^{42}$Sc cannot be described remarkably in the recently developed model~\cite{PhysRevC.109.064302}.
This seems to suggest that the neutron-proton correlation around Fermi surface is significantly underestimated in our calculations.
Intriguingly, $^{42}$Sc can be regarded as the combination of the two valence nucleons and the relatively stable core of $^{40}$Ca.
In order to declare the neutron-proton correlation around Fermi surface, we further redefine the three-point formula along isotonic chain as follows:
\begin{small}
\begin{eqnarray}\label{oef2}
\Delta(\Delta{R_{\mathrm{ch}}})(N,Z)=\frac{1}{2}[\Delta_{r}(N,Z)-\Delta_{r}(N,Z-1)-\Delta_{r}(N,Z+1)],\nonumber\\
\end{eqnarray}
\end{small}
where the values of $\Delta_{r}(N,Z)$ can be obtained through Eq.~(\ref{oef1}).

\begin{figure}
\resizebox{0.52\textwidth}{!}{%
  \includegraphics{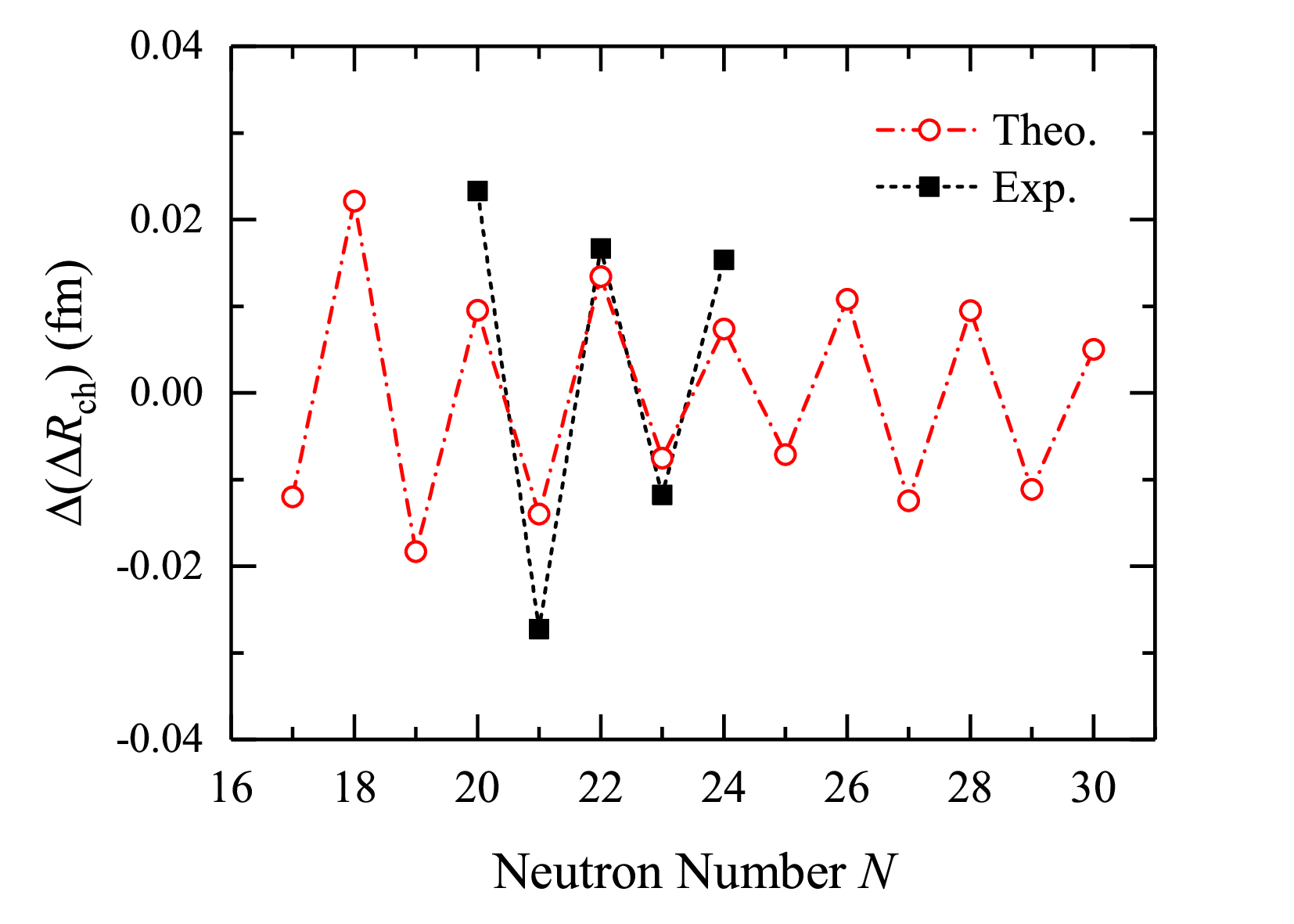}
}
\caption{(Color online) Evolution in $\Delta(\Delta{R_{\mathrm{ch}}})$ derived from Eq.~(\ref{oef2}) versus neutron numbers. Experimental data are taken from Refs.~\cite{ANGELI201369,LI2021101440,PhysRevLett.131.102501,Koszorus2020mgn}.}
\label{fig3}       
\end{figure}

In Fig.~\ref{fig3}, the values of $\Delta(\Delta{R_{\mathrm{ch}}})$ derived from Eq.~(\ref{oef2}) are used to measure the simultaneously unpaired neutron-proton correlation around Fermi surface.
From this figure, one can find that the OES in $\Delta(\Delta{R_{\mathrm{ch}}})$ can also be observed with the increasing neutron numbers. The calculated results falling at the neutron numbers $N=22$, $23$, and $24$ can reproduce the experimental data well.
At the neutron number $N=21$, the calculated result is apparently underestimated in comparison with the experiment.
This means that the simultaneously unpaired neutron-proton correlation around Fermi surface is underestimated significantly for $^{42}$Sc, namely the $\delta$ value in Eq.~(\ref{cp2}) is much lower.
The same scenario can also be observed at the neutron number $N=20$ due to the overestimated charge radii of $^{40,41}$Sc~\cite{PhysRevC.109.064302}.
Thus more reliable experimental data are urgently required in the proceeding work.

\begin{figure}
\resizebox{0.52\textwidth}{!}{%
  \includegraphics{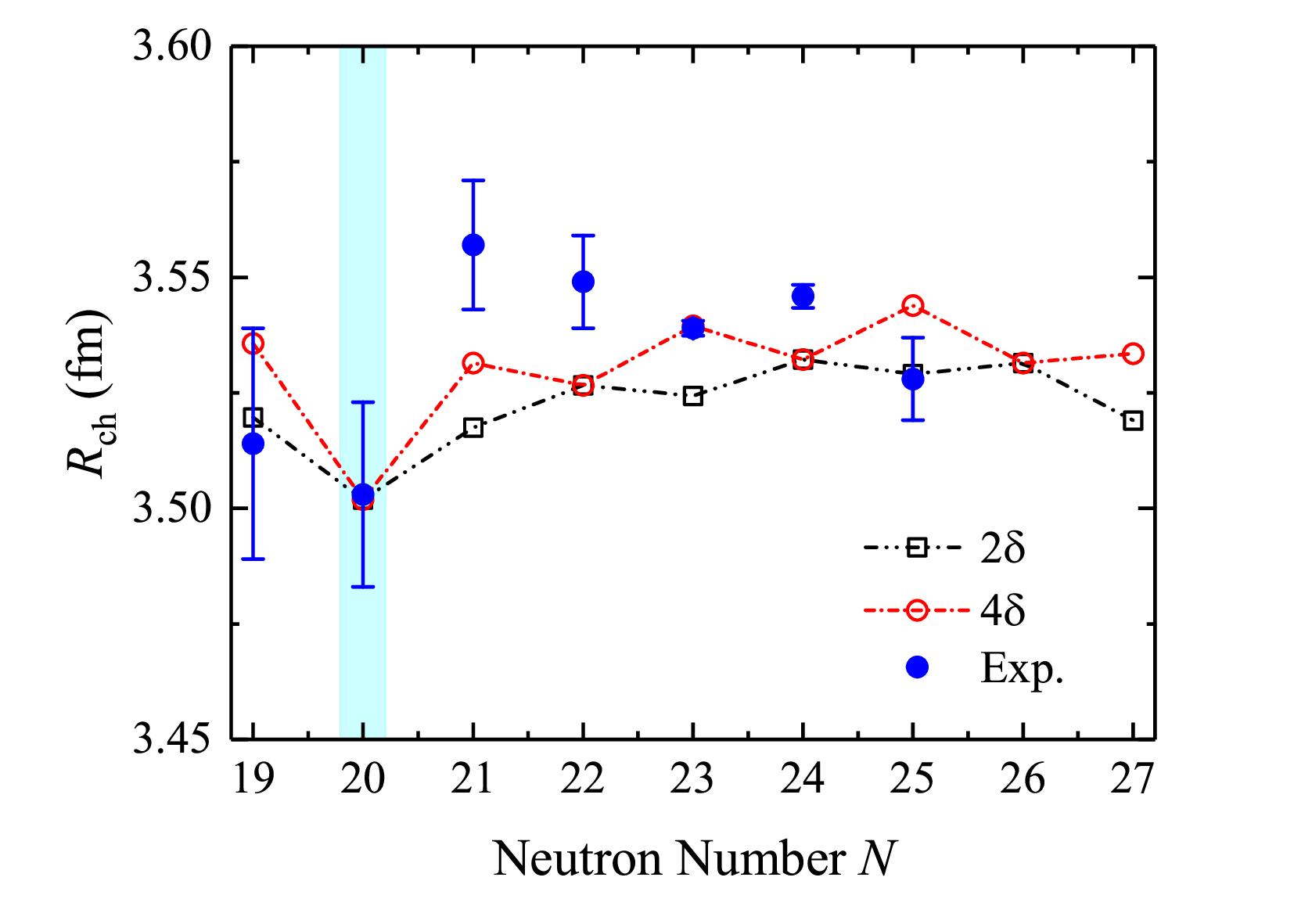}
}
\caption{(Color online) Charge radii $R_{\mathrm{ch}}$ of scandium isotopes derived from Eq.~(\ref{cp2}) with $2\delta$ and $4\delta$ modifications as a function of neutron numbers. Experimental data are taken from Refs.~\cite{ANGELI201369,LI2021101440,PhysRevLett.131.102501,Koszorus2020mgn}.}
\label{fig4}       
\end{figure}

Shown in Ref.~\cite{PhysRevC.109.064302}, charge radii of $^{40,41}$Sc isotopes can be reproduced well with the effective force PK1 in comparison to those obtained by NL3 set.
Therefore, the effective force PK1~\cite{PhysRevC.69.034319} is used to pick the trend of changes of charge radii along scandium isotopes in the following discussion.
As mentioned above, the neutron-proton correlation seems to be underestimated in Eq.~(\ref{cp2}). To facilitate the quantitative comparison, charge radii of the scandium isotopes are depicted as shown in Fig.~\ref{fig4}. In this figure, the last term in the  Eq.~(\ref{cp2}) is sequently replaced with the $2\delta/\sqrt{A}$ and $4\delta/\sqrt{A}$ modifications.
With increasing the neutron-proton correlations derived from the simultaneously unpaired neutron and proton, charge radius of $^{42}$Sc is actually enlarged. This leads to the significantly shrunken trend of charge radii at the neutron number $N=20$.
However, the trend of changes of charge radii cannot be described well around $N=24$.
This may result from the time-reversal symmetry breaking of the unpaired nucleons.
As shown in Fig.~\ref{fig4}, the value of charge radius for $^{42}$Sc is still underestimated.
This means more underlying mechanisms should not be considered appropriately.

Here one should be mentioned that the phenomenological term is used to capture the correlation between the simultaneously unpaired neutron and proton in Eq.~(\ref{cp2}). The calculated results show that this incorporated term cannot cover the enlarged change of charge radii from $^{41}$Sc to $^{42}$Sc adequately as shown in Fig.~\ref{fig4}. In addition, more underlying mechanisms in treating pairing correlations, such as particle number conservation~\cite{An_2020}, quadrupole deformation~\cite{PhysRevC.108.024310}, and the appropriate Bogoliubov transformation~\cite{PhysRevLett.77.3963,Meng1998}, should be taken into account. As demonstrated in Refs.~\cite{PhysRevC.84.024306,PhysRevC.91.027304}, beyond-mean-field approach plays an important role in describing the bulk properties of finite nuclei. As shown in Fig.~1~(c), the potential energy curve of $^{42}$Sc is relatively soft. This seems to suggest that the beyond-mean-field corrections should be considered in analyzing the enlarged charge radius of $^{42}$Sc.
\section{Summary}\label{Sec5}
The signature of nuclear shell structures is generally observed in nuclear charge radii at the traditional neutron magic numbers $N=28$, $50$, $82$, and $126$. However, as a participant of the traditional neutron magic number $N=20$, this profound kink structure is absent in the Ca, K, and Ar isotopic families. Interestingly, recent study gives the reliable values of charge radii for $^{40}$Sc and $^{41}$Sc isotopes by using the laser spectroscopic approach~\cite{PhysRevLett.131.102501}.
Combining the existing literature in Ref.~\cite{Avgoulea_2011}, the abrupt increase of charge radii can be observed significantly across $N=20$ shell closure. This unexpected trend provides an long-outstanding challenge in theoretical study.
This dues to the fact that $^{42}$Sc can be regarded as the formation of $^{40}$Ca core plus the simultaneously unpaired neutron and proton in the proton-proton and neutron-neutron channel, respectively.

This work focuses on the key aspect why the rapid raise of charge radii can be happened unexpectedly.
The recently developed model considering the simultaneously unpaired neutron-proton correlation around Fermi surface is employed in our discussion~\cite{PhysRevC.109.064302}. By analyzing various blocking configurations mixing of the simultaneously unpaired neutron and proton, it suggests that the influence of quadrupole deformation and blocking configurations on determining the increasing trend in charge radii of $^{42}$Sc should be considered properly.
Through the values of $\Delta(\Delta{R_{\mathrm{ch}}})$, we seem to infer that the neutron-proton correlation for the last unpaired nucleons in $^{42}$Sc is underestimated significantly. Meanwhile, the overestimated charge radii of $^{40,41}$Sc lead to the underestimated neutron-proton correlation at the neutron number $N=20$.
As shown in Ref.~\cite{MILLER2019360}, the overlap between the neutron and proton wave functions can be used to measure the neutron-proton correlation. This seems to provide a microscopic access to capture the correlation between the simultaneously unpaired neutron and proton.
Meanwhile, it should be mentioned that the isospin symmetry breaking effect can also have an influence on the determination of charge radii~\cite{PhysRevC.105.L021304,PhysRevC.106.L061306}.

As is well known, nuclear charge radii can be influenced by various mechanisms~\cite{PhysRevLett.128.152501}.
Reliable description of charge radii can be used to extract the information about proton size~\cite{ZHANG20241647}.
Highly linear correlation between the charge radii difference of mirror-pair nuclei and the slope parameter of symmetry energy has been built to pin down the isospin interactions in the equation of state of asymmetric nuclear matter~\cite{PhysRevLett.119.122502,PhysRevC.97.014314,PhysRevResearch.2.022035,XU2022137333,PhysRevLett.127.182503,PhysRevLett.130.032501,
nuclscitech34.119,NST35_182,PhysRevC.108.015802,PhysRevLett.132.162502}.
Therefore, the underlying mechanisms should be taken into account properly in describing the nuclear charge radii.
And more reliable charge radii data are expected in experiments.

\section{Acknowledgements}\label{ackn}
This work was partly supported by the Open Project of Guangxi Key Laboratory of Nuclear Physics and Nuclear Technology, No. NLK2023-05, the Central Government Guidance Funds for Local Scientific and Technological Development, China (No. Guike ZY22096024), the Natural Science Foundation of Ningxia Province, China (No. 2024AAC03015), and the Key Laboratory of Beam Technology of Ministry of Education, China (No. BEAM2024G04). X. J. was grateful for the support of the National Natural Science Foundation of China under Grants No. 11705118, No. 12175151, and the Major Project of the GuangDong Basic and Applied Basic Research Foundation (2021B0301030006).
L.-G. C. was grateful for the support of the National Natural Science Foundation of China under Grants No. 12275025 and No. 11975096, and the Fundamental Research Funds for the Central Universities (2020NTST06).
F.-S. Z. was supported by the National Key R$\&$D Program of China under Grant No. 2023YFA1606401 and the National Natural Science Foundation of China under Grants No. 12135004, No. 11635003, No. 11961141004, and No. 12047513.

%
%

\end{document}